\documentclass[aps,prc,floatfix,amsmath]{revtex4}
\usepackage{dcolumn}
\usepackage{amssymb}
\usepackage{mathrsfs}
\usepackage{supertabular}
\usepackage{color,graphicx}
\usepackage[bookmarksopen]{hyperref}
\newcommand{\Li}{\mathrm{Li}}

\begin{document}
\title{Non-Fermi liquid behavior of the drag and diffusion coefficients in QED plasma}
\author{Sreemoyee Sarkar}
\email{sreemoyee.sarkar@saha.ac.in}

\author{Abhee K. Dutt-Mazumder}
\email{abhee.dm@saha.ac.in}

\affiliation{High Energy Nuclear and Particle Physics Division, Saha Institute of Nuclear Physics,
1/AF Bidhannagar, Kolkata-700 064, INDIA}

\medskip

\begin{abstract}
We calculate the drag and diffusion coefficients in low temperature QED plasma and go beyond the leading order approximation. 
The non-Fermi-liquid behavior of these coefficients are clearly revealed. We observe that the subleading contributions 
due to the exchange of soft transverse photon in both cases are larger
than the leading order terms coming from the longitudinal sector. The results are presented
in closed form at zero and low temperature.
\end{abstract}

\maketitle

\section{Introduction}
It has been known for quite sometime now that a fermionic system interacting via 
the exchange of transverse gauge bosons exhibit deviations from the normal Fermi liquid behavior.
Such a characteristic feature, in presence of transverse or magnetic interactions, for the first time was 
reported in \cite{Holstein73} where
the specific heat of a degenerate electron gas was shown to contain correction terms involving
$\alpha_s T {\ln} T^{-1}$. This was interpreted to be a consequence of the long range behavior
of the magnetic interaction due to the absence of magnetostatic screening. Initially, such corrections were 
considered to be of little practical importance as, such  tiny an effect, was 
not likely to be detected experimentally. A decade later, however, the scenario 
changed and such investigations started attracting attentions in the context of strongly correlated electron system 
in which the gauge coupling is not the fine structure constant (1/137) but of order unity \cite{Chakravarty95}. Further impetus to these studies now come from another domain involving relativistic quark (or electron) gas at 
high density and zero or low temperature where the specific heat also contain such anomalous terms.
This seems to have serious implications in determining the thermodynamic and transport 
properties of the 
quark component of neutron or proto neutron stars, {\em viz.} entropy, pressure, 
specific heat, viscosity
etc \cite{Boyanovsky01,Ipp04, Gerhold204, Heiselberg93}. Similar non-Fermi-liquid terms, for ungapped quark matter, also appear 
in the calculation of neutrino emissivity and its mean free path \cite{Schafer04b,Pal11}. 
For quarks in a color superconducting state also, the chromomagnetic interaction
strongly influences the magnitude of the gap as shown in \cite{Brown00,Wang02}.

It is well known that the magnetic interaction in non-relativistic systems is suppressed in powers of $(v/c)^2$. The
scenario, however, changes as one enters into the relativistic domain, where it becomes 
important. Hence, in dealing with relativistic plasma one has to retain both electric and magnetic 
interactions mediated by the exchange of longitudinal and transverse gauge bosons like photons or
gluons. More interestingly, it is observed that for ultradegenerate case, both in QCD and QED, the transverse 
interactions not only become important but it dominates over its longitudinal counterpart; a characteristic behavior 
having a non-trivial origin residing in the analytical structure of the Fermion-self energy close to the Fermi
surface. This has been beautifully exposed in \cite{Gerhold05} where the authors calculate Fermionic dispersion relations 
in ultradegenerate relativistic plasmas and show how such non-Fermi liquid behavior emerges from the vanishing of the 
Fermion propagator near the Fermi surface by calculating the group velocity of the corresponding quasiparticle excitations.
One can also see, how the fractional power appears there \cite{Gerhold05}, due to the exchange of soft transverse gauge
boson in the small temperature expansion of the Fermion self-energy in ultradegenerate plasma similar to what
one encounters in the expansion of the thermodynamic potential or $C_v$ \cite{Boyanovsky01,Ipp04,Gerhold204}. Actually, the fermion self-energy close to 
the Fermi surface receives a logarithmic enhancement due to the exchange of magnetic gluons \cite{Schafer04}, 
this, in turn, leads to such dominance. A more rigorous discussion on how and why the dynamics changes near 
the Fermi surface leading to the break down of Fermi liquid behavior or vanishing of the step discontinuity  
can be found in \cite{Boyanovsky01b}. Departure from the Fermi liquid behavior has also been witnessed in the 
calculation of quasiparticle damping rate in ultradegenerate relativistic plasma \cite{Bellac97, Manuel00, Vanderheyden97}.
  
 The low temperature, high density region,
 commonly known as ultradegenerate plasma, is much less explored in comparison with the high temperature
 low density domain. In particular in the present work, we calculate fermionic drag ($\eta$) and longitudinal diffusion coefficients (${\cal B}_\| \ $) in this
regime and eventually extend it to the limiting case of zero temperature.  The salient feature here has been the
inclusion of the higher order terms both in the transverse and longitudinal sector with 
implications to be discussed later. The evaluation of drag (diffusion) coefficient is very similar to the damping
rate calculation with one difference {\em i.e.} here we weight the imaginary part of the self energy with
the energy (square momentum) transfer per scattering to obtain the desired result. Such calculations, as is
well known, are plagued with infrared divergences. There are well established techniques to handle such 
divergences both at finite and zero temperature where one divides the interactions into two regions one involving
the exchange of soft gauge bosons while the other involves hard momentum transfer \cite{Braaten91}. For the former one uses the bare photon (gluon) propagator
and for the latter the hard thermal/density loop (HTL/HDL) resummed propagator is used. One interesting departure from
the high temperature that is observed in dealing with plasma close to zero temperature is the following: in a hot plasma
both the hard and the soft part of the electric and magnetic interactions contribute at  same order of 
the coupling parameter. In the ultradegenerate plasma, or when the temperature is much smaller compared to 
the chemical potential, it is seen that the hard sector contribution come with higher order coupling
parameters than the soft sector. Even within the soft sector, for the longitudinal and transverse part, the
coupling parameter appears with different
powers \cite{Sarkar10}. 

The drag coefficient, as we know, is related to the energy loss suffered by the propagating particle in a plasma. This has
been studied extensively in a series of works for the last two decades 
\cite{Svetitsky88, Braaten191, Braaten291, Mazumder05,  Moore05, Roy06, Mustafa05, Peigne108, Peigne208, 
Beraudo06}. 
There also exist many calculations for the diffusion coefficients both for quantum electro and chromomagnetic plasma 
\cite{Svetitsky88, Moore05, Roy06, Beraudo06}. 
All these calculations are performed in situations where the temperature is high but the
chemical potential is zero except in \cite{Vija95,Das10} where numerical estimates of the energy loss or
drag and diffusion coefficients at non-zero chemical potential have been presented. There, to the best of our knowledge, 
exists only one calculation so far \cite{Sarkar10}, where the analytical results for $\eta$ and  (${\cal B}_\| \ $) for ultradegenerate relativistic plasma 
have been presented. There, we have restricted ourselves only to the leading order results and have shown that the 
drag and diffusion coefficients are dominated by the soft transverse photon exchanges while the longitudinal terms
are subleading. Here, we go beyond the leading order and reveal the importance of the subleading terms in the
transverse sector. The approach we adopt in this work is, however, different from the previous one and more in line with
\cite{Gerhold05}. The connections, nevertheless, are made at appropriate places. Here, we probably should mention that
 the dominance of next to leading order (NLO) transverse term over the longitudinal one does not imply breakdown of the perturbation
 series. Because, the next to leading order terms in transverse or longitudinal sector individually are smaller than the corresponding
 leading parts.      

The plan of this paper is as follows: in section II the formalism is set forth. In subsection A, we evaluate
the drag coefficient in the domain low temperature and eventually arrive at the zero temperature results
by taking the appropriate limit. In the next subsection we present the results for the diffusion coefficient
both at zero and small temperature. In section III we conclude. 

\section{Formalism}

\begin{figure}[htb]
\begin{center}
\resizebox{12.5cm}{4.75cm}{\includegraphics{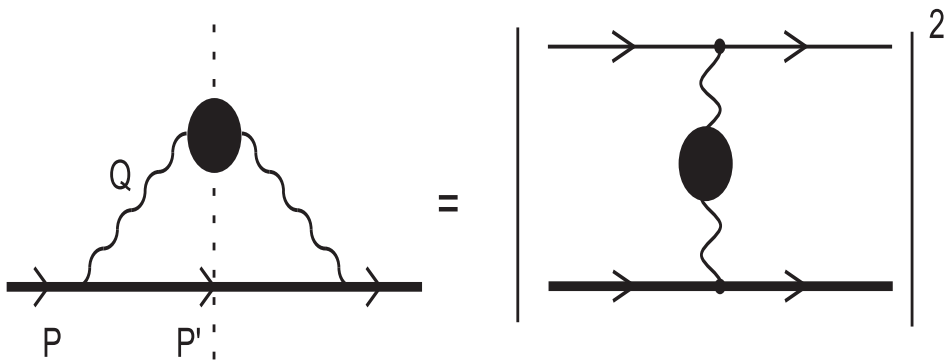}}
\caption{Fermion-fermion scattering with screened interaction.
\label{fig1}}
\end{center}
\end{figure}
The drag coefficient of a quasiparticle having energy $(E)$ is incidentally related to the energy loss of the propagating particle which undergoes collisions with 
the constituents of the plasma {\em viz.} the electrons: 
\begin{eqnarray}
\eta&=& {1\over E }\int d\Gamma \omega,
\label{eta}
\end{eqnarray}
$d\Gamma$ is the differential interaction rate \cite{Bellac}.
This expression is quite general and can be used to calculate collisional energy loss both for the finite temperature
and/or density. The phase space will be different
due to the modifications of the distribution functions depending upon
the values of $\mu$ and $T$. The imaginary part of fermion self
energy diagram basically gives the the damping rate of a hard fermion. This damping mechanism is equivalent to elastic scattering off the thermal electrons via the exchange
of a collective photon,

\begin{equation}
 \Gamma(E)=-{1\over 2 E}  {\rm Tr}\,\left[{\rm Im}\,\Sigma(p_0+i\epsilon\ ,{\bf
p}){P\llap{/\kern1pt}}\right] \Big|_{p_0=E}  \ .
\label{gamma}
\end{equation}
The full fermion self-energy represented in Fig.(\ref{fig1}) can be written explicitly 
as: 
\begin{equation}
\Sigma(P)= e^2T\sum_s\int{{\rm d}^3q\over (2\pi)^3}\gamma_\mu \,
S_f(i(\omega_n-\omega_s),{\bf p-q})\gamma_\nu \,
\Delta_{\mu\nu}(i\omega_s,{\bf
q}) \ ,
\label{sigma}
\end{equation}

where, $p_0=i\omega_n+\mu$, $q_0=i\omega_s$. $\omega_n=\pi(2 n+1)T$ and 
$\omega_s=2\pi s T$ are the Matsubara frequencies for fermion and boson 
respectively with integers $n$ and $s$. After performing the sum over Matsubara frequency in Eq.(\ref{sigma}), $i\omega_n+\mu$ is analytically continued 
to the Minkowski space $i\omega_n+\mu\rightarrow p_0+i\epsilon$, with 
$\epsilon\rightarrow0$. The blob in the wavy line of Fig.(\ref{fig1}) represents HTL/HDL corrected photon propagator which is in the 
Coulomb gauge is given by \cite{Bellac, Kapusta},

\begin{eqnarray}
\Delta_{\mu\nu}(Q)=\delta_{\mu 0}\delta_{\nu 0} \, \Delta_l(Q)
+{P}^t_{\mu \nu}\Delta_t(Q) \ ,
\label{delta}
\end{eqnarray}

with, ${P}^t _{i j} = (\delta_{ij}-\hat q_i\hat q_j), {\hat q}^i = {\bf q}^i/|{\bf q}|$, 
  ${P}^t_{i0} = {P}^t_{0i} =
{P}^t_{00}=0$ and $\Delta_{l}$, $\Delta_{t}$ are given by \cite{Bellac,Kapusta},
\begin{eqnarray}
\Delta_{l} (q_0, q)   & = & {-1\over q^2+\Pi _{l}},  \\
 \Delta_{t}  (q_0, q) &  =  & {-1\over q_0^2-q^2-\Pi _{t}} .
\label{delta_l_t}
\end{eqnarray}
Here we introduce the spectral functions $\rho_{l,t}$ as\cite{Bellac,Kapusta}:
 
\begin{equation}
\frac{\rho_{l,t} (q_0,q)}{2 \pi} = Z_{l,t} 
\left[\delta(q_0 - \omega_{l,t} (q)) -
\delta(q_0 + \omega_{l,t} (q)) \right] + \beta_{l,t} (q_0,q) \ .
\end{equation}
The poles $\omega_{l,t}$  are  the solutions of the dispersion relations. The $\delta$ function corresponds to the (time-like) poles
 of the resumed propagator and $\beta_{l,t}$ represent cuts.
The latter terms {\em i.e} Landau damping pieces of the spectral functions
 are non-vanishing only
for $q_0^2 \leq q^2 $ and are given by,
\begin{mathletters}
\label{betas}
\begin{eqnarray}
\beta_l(q_0,q) & =&
 \frac{m_D^2\,  x \,\Theta(1-x^2)}{2
\left[ q^2 +m_D^2  \left( 1 - \frac{x}{2} 
\ln {\Big| \frac{x+1}{x-1} \Big|} \right) \right]^2 + \frac{m_D^4  \pi^2 x^2 }{4}
} \ , \nonumber\\
\beta_t(q_0,q) & =&
 \frac{m_D^2\, \, x \,(1-x^2)\Theta(1-x^2)}{
\left[2 q^2(x^2 -1) -m_D^2 x^2  \left( 1 + \frac{(1-x^2)}{2x} 
\ln {\Big| \frac{x+1}{x-1} \Big|} \right) \right]^2 + \frac{m_D^4  \pi^2 x^2 (1- x^2)^2}{4}} \ ,
\label{beta_l_t}
\end{eqnarray}
\end{mathletters}
where $x =q_0/ q$. The Debye
mass is $m_D^2= {e^2 \over \pi^2}\Big(\mu^2+{\pi^2T^2\over 3}\Big)$.


At the leading order these are derived from the one-loop photon self-energy where the loop momenta 
are assumed to be hard in comparison to the photon momentum \cite{Bellac,Kapusta}. In the literature the formalism is known as the HTL/HDL 
approximation as discussed in \cite{Bellac, Kapusta, Manuel96, Manuel00, Pal10}.

In Eq.(\ref{sigma}), fermion propagator has the following spectral representation with the notation 
$\bf{k}=(\bf{p}-\bf{q})$ \cite{Bellac},
\begin{equation}
S_f(i\omega_n,{\bf k})=\int_{-\infty}^\infty{{\rm d}k_0\over
2\pi}\,{{K\llap{/\kern3pt}}\rho_f(K)\over
k_0-i\omega_n-\mu} \ .
\label{ferm_spec_func}
\end{equation}

Taking the imaginary part of Eq.(\ref{sigma}), the scattering rate with the help of Eq.(\ref{gamma}) can be calculated.
 One then inserts the energy exchange $\omega$ in the expression of 
$\Gamma$ and calculate $\eta$ from Eq.(\ref{eta}), to obtain,

\begin{eqnarray}
 \eta&=&{\pi e^{2}\over E^2} \int{{\rm d}^3q\over
(2\pi)^3}\int_{-\infty}^\infty{{\rm d}k_0\over 2\pi}\rho_f(k_0)
\int_{-\infty}^\infty{{\rm d}q_0\over 2\pi}q_{0} \nonumber\\
&\times &
(1+ n(q_0)- {\bar n}(k_0))  \delta(E-k_0-q_0)\nonumber \\
&\times &[p_0k_0+{\bf p\cdot
k}] 
  \rho_{l}(q_0,q) \nonumber \\
&+&2[p_0k_0-({\bf p\cdot \hat q})({\bf k\cdot\hat
q})]\rho_{t}(q_0,q). 
\label{eta_def}
\end{eqnarray}
 The energy conserving delta function in the last equation deletes the contribution from the delta function and therefore
 $\eta$ recieves contribution only from the cuts. The same holds true for diffusion coefficients ${\cal B}_{\| \ , \perp}$ as well. 
 In the above equation $n$ and 
${\bar n}$ are the Bose-Einstein and the Fermi-Dirac
distribution functions:
\begin{equation}
n(q_0)={1\over {\rm e}^{\beta q_0}-1} \ , \qquad
{\bar n} (k_0)={1\over {\rm e}^{\beta (k_0-\mu)}+1} \ ,
\label{dist_func}
\end{equation}
where, $\beta={1\over T}$. Eq.(\ref{eta_def}) is the general expression of drag coefficient.

Apart from $\eta$, the quantity momentum diffusion coefficient ($B_{ij}$), could be of importance in the study of fermion 
propagating in the plasma \cite{Svetitsky88, Moore05, Roy06, Beraudo06}. It can be defined as follows \cite{Svetitsky88,Moore05,Roy06,Beraudo06},
\begin{eqnarray}
 { B_{ij}}=\int d\Gamma q_i q_j.
\label{diff_1}
\end{eqnarray}
Decomposing $B_{ij}$ into longitudinal ($B_ {\| \ }$) and transverse components ($B_\perp$) we get the following expression,
\begin{eqnarray}
 B_{ij}=B_\perp(\delta_{ij}-\frac{p_i p_j}{p^2}) + B_{\| \ } \frac{p_ip_j}{p^2}.
\label{diff_2}
\end{eqnarray}

The imaginary part of the Eq.(\ref{sigma}) multiplied by the square of the longitudinal momentum transfer in the fermion fermion scattering
 gives the expression for ${\cal B}_\| \ $. Using Eqs.(\ref{diff_1}) and (\ref{diff_2}), longitudinal momentum diffusion 
coefficient ($ B_{||}={\cal B}$) can be written as follows,
\begin{eqnarray}
 {\cal B}&=&{\pi e^{2}\over E} \int{{\rm d}^3q\over
(2\pi)^3}\int_{-\infty}^\infty{{\rm d}k_0\over 2\pi}\rho_f(k_0)
\int_{-\infty}^\infty{{\rm d}q_0\over 2\pi}q_{\| \ }^2 \nonumber\\
&\times &
(1+ n(q_0)- {\bar n}(k_0))  \delta(E-k_0-q_0)\nonumber \\
&\times &[p_0k_0+{\bf p\cdot
k}] 
  \rho_l(q_0,q) \nonumber \\
&+&2[p_0k_0-({\bf p\cdot \hat q})({\bf k\cdot\hat
q})]\rho_t(q_0,q). 
\label{diff_exp}
\end{eqnarray}
Here, $q_{\| \ }=q {\rm cos} \theta $ {\em i.e} the longitudinal momentum transfer. 

\subsection{Drag coefficient when $|E-\mu|$ $\sim$ T}

In this section we calculate the drag coefficient ($\eta$)  when $T\sim |E-\mu|\ll e\mu\ll\mu$, this is the region
which is relevant for the astrophysical applications. It has been mentioned already that evaluation of $\eta$ is
plagued with infrared divergences. To circumvent this problem, as mentioned in the introduction, the region of
integration as it appears below has to be divided into two regions distinguished by the scale of
the momentum transfer {\em i.e.} the soft and the hard sector. For the former, we use the one loop resummed
propagator with a finite upper limit on the momentum which is designated as $q^*$ and for the latter we use
the bare photon propagator. Following this prescription, for the soft part, one writes:
\begin{eqnarray}
\eta\Big|^{\rm soft}(E) 
& \simeq & {e^2\over 8 \pi^2E} \int_0^{q^*} {\rm d}q q^3 \int_{-1}^1{\rm d}xx 
( 1+n(qx) \, -\bar n(E-\mu-qx) ) \,\nonumber \\ 
&  \times &\{ \rho_l(qx, q)
+(1-x^2)\rho_t(qx,q) \}\nonumber\\
\label{eta_def}
\end{eqnarray}
From the expression after subtracting the energy independent part we have \cite{Gerhold05},
\begin{eqnarray}
\eta\Big|^{\rm soft}(E)-\eta\Big|^{\rm soft}_{E=\mu}&=&-{e^2\over8\pi^2 E}\int_0^{q^*} dq\,q^3\int_{-1}^1dxx
  \left(\bar n(E-\mu-qx)-\bar n(-qx)\right)\nonumber\\
  &&
\times\left[(1-x^2)\rho_t(qx,q)+\rho_l(qx,q)\right]. \label{eta_exp}
\end{eqnarray}

First, we calculate the transverse photon contribution then the longitudinal one. For this in Eq.(\ref{eta_exp}) we substitute $q$ and $q_0$ by introducing dimensionles variables $z$ and $v$, 
\begin{equation}
  q=2q_sz/(\pi v)^{1/3}, \quad q_0=Tv, \label{su1}
\end{equation}
where, $q_s$ is the screening distance in the magnetic sector, and we take $a={T\over m_D}\ll 1$. From the above substitutions it immediately
 follows that,
\begin{eqnarray}
 q=m_D a^{1/3}z,\quad x=a^{2/3}v/z.\label{su2}
\end{eqnarray}
 
After expanding the
 integrand with respect to $a$ we find for the transverse contribution of $\eta$,

\begin{eqnarray}
  &&\!\!\!\!\!\!\!\!\!\!\!
\eta\Big|^{\rm soft}_t(E)-\eta\Big|^{\rm soft}_{t, E=\mu}
  =-{e^2m_D^2a^2\over2\pi E}\int_{-{q^*\over am_D}}^{q^*\over am_D}dvv 
  \int_{a^{2/3}|v|}^{q^*\over a^{1/3}m_D}dz\,{e^\alpha-1\over(1+e^v)(1+e^{\alpha-v})}\nonumber\\
  &&\qquad\times\bigg[-{z^2v\over v^2\pi^2+4z^6}+{16v^3z^4\over(v^2\pi^2+4z^6)^2}a^{2/3}
  +{16v^5(v^2\pi^2-12z^6)\over(v^2\pi^2+4z^6)^3}a^{4/3}+\ldots\bigg],\nonumber\\
\label{eta_t_1}
\end{eqnarray}
where, $\alpha={|E-\mu|\over T}\sim O(1)$. Here, we neglect the terms which are more than $a^{10\over 3}$ and $({m_D\over q^*})^4$. After $z$ integration we obtain,
\begin{eqnarray}
\eta\Big|_t^{soft}(E)-\eta\Big|^{\rm soft}_{t, E=\mu}={e^2m_D^2\over E}\int_{-\infty}^\infty dv{e^\alpha-1\over(1+e^v)(1+e^{\alpha-v})}\left({va^2\over 24\pi}-{2^{1\over3}v^{5\over3}a^{8\over3}\over 9 \pi^{7\over3}}
-{20\times 2^{2\over3}v^{7\over3}a^{10\over3}\over 27 \pi^{11\over3}}+\ldots\right).
\label{eta_t_2}
\end{eqnarray}

Now, we use the formula for $v$ integration sending the integration limits to $\pm \infty$,
\begin{eqnarray}
  &&\!\!\!\!\!\!\!\!\!
  \int_{-\infty}^\infty dv{e^\alpha-1\over(1+e^v)(1+e^{\alpha-v})}
  |v|^\lambda\nonumber\\
  &&\!\!\!\!=\Gamma(\lambda+1)\left[\Li_{\lambda+1}(-e^{-\alpha})-\Li_{\lambda+1}(-e^{\alpha})\right].
  \quad\forall \lambda\ge0 \nonumber\\
\label{form}
\end{eqnarray}

Clearly the expression for $\eta\Big|_t^{soft}$ is Polylogarithmic in nature, 
\begin{eqnarray}
\eta\Big|_t^{soft}(E)-\eta\Big|^{\rm soft}_{t, E=\mu}&=&{e^2m_D^2\over E}\Bigg\{{a^2\over 24\pi}\left[\Gamma(\textstyle{2})
  \left(\Li_{2}(-e^{-\alpha})+\Li_{2}(-e^\alpha)\right)\right]\nonumber\\
&-&{2^{1/3}a^{8/3}\over9\pi^{7/3}}\left[\Gamma(\textstyle{8\over3})
  \left(\Li_{8/3}(-e^{-\alpha})+\Li_{8/3}(-e^\alpha)\right)\right]\nonumber\\
&-&{20\times2^{2/3}a^{10/3}\over9\pi^{11/3}}\left[\Gamma(\textstyle{10\over3})
 \left(\Li_{10/3}(-e^{-\alpha})+\Li_{10/3}(-e^\alpha)\right)\right]
+\ldots\Bigg\}. 
\label{eta_t_3}
\end{eqnarray}
The above expression can be written in the following form,
\begin{eqnarray}
\eta\Big|_t^{soft}(E)-\eta\Big|^{\rm soft}_{t, E=\mu}&=&{e^2m_D^2\over E}\Bigg\{{1\over 48\pi}
\left({T\over m_D}h_1\left({(E-\mu)\over T}\right)\right)^2-{3\times2^{1/3}\over 72\pi^{7/3}}\left({T\over m_D}h_2\left({(E-\mu)\over T}\right)\right)^{8\over 3}\nonumber\\
&-&{6\times2^{2/3}\over9\pi^{11/3}}\left({T\over m_D}h_3\left({(E-\mu)\over T}\right)\right)^{10\over 3}\Bigg\} ,
\label{eta_t_4}
\end{eqnarray}
where,
\begin{eqnarray}
h_1\left({(E-\mu)\over T}\right)&=&\left[\Gamma(\textstyle 3)
  \left(\Li_{2}(-e^{-\alpha})-\Li_{2}(-e^\alpha)\right)\right]^{1\over 2}\nonumber\\
h_2\left({(E-\mu)\over T}\right)&=&\left[\Gamma\left(\textstyle{11\over3}\right)
  \left(\Li_{8/3}(-e^{-\alpha})-\Li_{8/3}(-e^\alpha)\right)\right]^{3/8}\nonumber\\ 
h_3\left({(E-\mu)\over T}\right)&=&\left[\Gamma\left(\textstyle{13\over3}\right)
  \left(\Li_{10/3}(-e^{-\alpha})-\Li_{10/3}(-e^\alpha)\right)\right]^{3/10}. 
\label{eta_t_5}
\end{eqnarray}
From the expression (\ref{eta_t_3}) it is evident that the expression contains fractional powers in $(E-\mu)$. This nature is basically
a non-Fermi-liquid behavior of ultradegenerate relativistic plasma.
After the magnetic part we derive the expression
 of the electric part. 

In case of the electric part we substitute $q=q_s y$ and $q_0=Tu/y$, or $q=m_Dy$ and $x=au/y$. Though the substitutions in electric and magnetic
 sectors look different, but the nature of substitutions can be seen from the structure of $\beta_{l,t}$ (Eq.(\ref{beta_l_t})). 
As screening length is different in electric and magnetic sectors the substitutions therefore involve different coefficients 
of $m_D$ and $T$ for the transverse and the longitudinal case \cite{Gerhold05}. The longitudinal term after simplification like transverse
 one becomes,
\begin{eqnarray}
\eta\Big|_l^{soft}(E)-\eta\Big|^{\rm soft}_{l, E=\mu}&=&{e^2m_D^2 a^3\over 32E}\left[\Gamma(\textstyle 3)
  \left(\Li_{3}(-e^{-\alpha})-\Li_{3}(-e^\alpha)\right)\right]+O(a^4).
\label{eta_l_2}
\end{eqnarray}
Again for the leading term we can write it in the following form,
\begin{eqnarray}
\eta\Big|_l^{soft}(E)-\eta\Big|^{\rm soft}_{l, E=\mu}&=&{e^2m_D^2 \over  96 E}\left({T\over m_D}g_1\left({(E-\mu)\over T}\right)\right)^{3},
\label{eta_l_4}
\end{eqnarray}
where,
\begin{eqnarray}
g_1\left({(E-\mu)\over T}\right)=\left[\Gamma(\textstyle 4)
  \left(\Li_{3}(-e^{-\alpha})-\Li_{3}(-e^\alpha)\right)\right]^{1\over3}.
\label{func}
\end{eqnarray}
The final expression for drag-coefficient then becomes,
\begin{eqnarray}
\eta&=&{e^2m_D^2\over E}\Bigg\{{1\over 48\pi}
\left({T\over m_D}h_1\left({(E-\mu)\over T}\right)\right)^2-{3\times2^{1/3}\over 72\pi^{7/3}}\left({T\over m_D}h_2\left({(E-\mu)\over T}\right)\right)^{8\over 3}\nonumber\\
&-&{6\times2^{2/3}\over9\pi^{11/3}}\left({T\over m_D}h_3\left({(E-\mu)\over T}\right)\right)^{10\over 3}\Bigg\}
+{e^2m_D^2 \over  96E}\left({T\over m_D}g_1\left({(E-\mu)\over T}\right)\right)^{3}.
\label{eta_full_exp}
\end{eqnarray}
In the zero temperature limit the functions behave as $h_i(\alpha)\rightarrow|\alpha|$ and $g_i(\alpha)\rightarrow|\alpha|$. Hence,
 $\eta$ in the extreme zero temperature limit becomes,
\begin{eqnarray}
\eta={e^2 |E-\mu|^2\over 48\pi E}-{3\times2^{1/3}e^2m_D^2\over 72\pi^{7/3}E}\left({|E-\mu|\over m_D}\right)^{8\over 3}+{e^2|E-\mu|^{3} \over 96m_DE}+\cdots.
\label{eta_final}
\end{eqnarray}
This is the result for zero temperature plasma. Both the first and the second term here come from the transverse sector while
the last piece emanates from the longitudinal interactions. The appearance of the second term with fractional power both
in Eqs.(\ref{eta_full_exp}) and (\ref{eta_final}) clearly show that full contributions to $\eta$ cannot be obtained by adding
leading order contributions of the transverse and longitudinal photon exchange as the subleading terms of the former is
larger than the leading order contribution of the later. This observation, in connection to the evaluation of Fermion
self energy was first noted in \cite{Gerhold05} and was overlooked in
\cite{Bellac97,Vanderheyden97,Manuel00,Sarkar10}. The zero temperature leading order contributions for $l$ and $t$ part are, however,
 consistent with our previous calculation reported in \cite{Sarkar10}. It is
needless to mention here that such characteristic feature, also known as non-Fermi liquid behavior, 
can be attributed to the absence of the magnetostatic screening as noted in the introduction. 

\begin{figure}
\begin{center}
\resizebox{7.5cm}{5.75cm}{\includegraphics{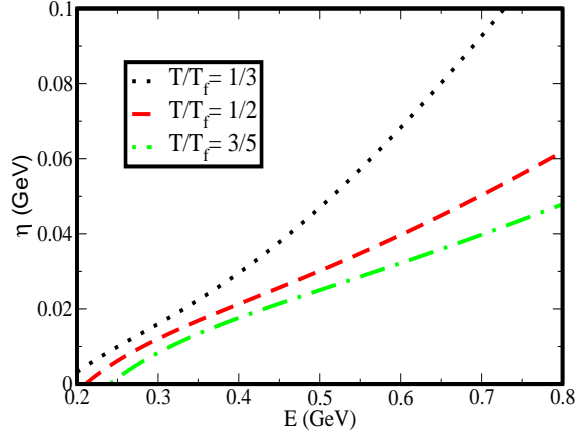}}
\end{center}

\caption{(Color online) The next to leading order drag coefficient at $T/T_f=1/3$ (dotted curve), $T/T_f= 1/2$ (dashed
 curve), $T/T_f=3/5$ (dash dotted curve).}
\label{nlo_eta}
\end{figure}
 In Fig.(\ref{nlo_eta}) we have plotted $\eta$ versus energy of the incoming fermion in the small temperature 
$(T/T_f<<1)$ region where $T_f=\mu/k_B$ is the Fermi temperature. From the figure it is evident that with increasing
 $T/T_{f}$, $\eta$ decreases. This trend is consistent with what one finds for 
the fermionic damping rate at small temperature \cite{Gerhold05}.   

 So far we have not discussed about the hard sector and tacitly assumed that the entire contribution to $\eta$ in
the relevant domains {\em i.e.} for small and zero temperature, come from the soft photon exchange. This, for
degenerate plasma, is indeed so, as demonstrated explicitly in \cite{Sarkar10}. In \cite{Sarkar10}, it was shown that the leading order of the hard sector fails to contribute to  
$\eta$ at least up to  $O(e^2)$. As, in the present work, on the other hand
we go beyond the leading order, in principle, one should calculate the NLO part for the hard sector as well
and see if the hard sector contributes to the drag and diffusion co-efficients in this case. But such an 
explicit calculation has not been done here. One can justify this 
omission on the ground that, for the soft sector, we see  even after the inclusion of the NLO corrections
no intermediate cut-off ($q^*$) dependent term appear up to $O(e^2)$. Therefore, in the spirit of our previous
work \cite{Sarkar10}, we conclude up to this order the entire contribution comes 
from the soft sector providing indirect justification of this omission. The finite temperature NLO calculation can shed 
further light on this issue \cite{Aurenche99,Carrington07,Carrington08}.
   

\subsection{Diffusion coefficient when $|E-\mu|\simeq T$}


Along with $\eta$, momentum diffusion coefficient ($B_{ij}$), \cite{Svetitsky88, Moore05, Roy06, Beraudo06} is another relevant quantity to study the equilibration of a fermion propagating in the plasma. 
For Coulomb plasma $\eta$ and the longitudinal momentum 
diffusion coefficient (${\cal B}$) are related {\em via} Einstein's Relation (ER). In this section we study the nature of longitudinal diffusion
coefficient in the low temperature region. In the soft region the expression looks like,
\begin{eqnarray}
{\cal B}\Big|^{\rm soft}(E) 
& \simeq & {e^2\over 8 \pi^2} \int_0^{q^*} {\rm d}q q^4 \int_{-1}^1{\rm d}xx^2 
( 1+n(qx) \, -\bar n(E-\mu-qx) ) \,\nonumber \\ 
&  \times &\{ \rho_l(qx, q)
+(1-x^2)\rho_t(qx,q) \}.\nonumber\\
\label{def_diff}
\end{eqnarray}

\begin{figure}
\begin{center}
\resizebox{7.5cm}{5.75cm}{\includegraphics{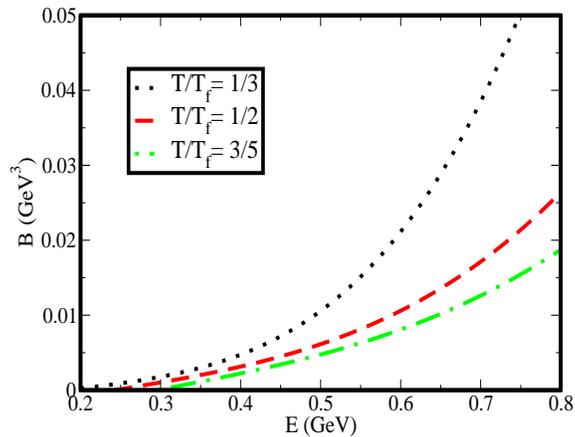}}
\end{center}
\caption{(Color online) The next to leading order diffusion coefficient at $T/T_f=1/3$ (dotted curve), $T/T_f= 1/2$ (dashed
 curve), $T/T_f=3/5$ (dash dotted curve).}
\label{nlo_diff}
\end{figure}

First, we calculate the transverse photon contribution then the longitudinal one. For the transverse photon propagator we 
proceed along the same line of the previous subsection and find,

\begin{eqnarray}
{\cal B}\Big|_t^{soft}(E)-{\cal B}\Big|^{\rm soft}_{t, E=\mu}&=&e^2m_D^3\Bigg\{{a^3\over 24\pi}\left[\Gamma(\textstyle{3})
  \left(\Li_{3}(-e^{-\alpha})-\Li_{3}(-e^\alpha)\right)\right]\nonumber\\
 &-&{2^{1/3}a^{11/3}\over9\pi^{7/3}}\left(\Gamma\left(\textstyle{11\over3}\right)
   (\Li_{11/3}(-e^{-\alpha})-\Li_{11/3}(-e^\alpha)
\right)\nonumber\\
&-&{20\times2^{2/3}a^{13/3}\over9\pi^{11/3}}\left(\Gamma\left(\textstyle{13\over3}\right)
  (\Li_{13/3}(-e^{-\alpha})-\Li_{13/3}(-e^\alpha)\right)+\ldots\Bigg\}.
\label{diff_t_1}
\end{eqnarray}
The above expression is written in the following form,
\begin{eqnarray}
{\cal B}\Big|_t^{soft}(E)-{\cal B}\Big|^{\rm soft}_{t, E=\mu}&=&e^2m_D^3\Bigg\{{1\over 72\pi}
\left({T\over m_D}h_4\left({(E-\mu)\over T}\right)\right)^3-{3\times2^{1/3}\over 99\pi^{7/3}}\left({T\over m_D}h_5\left({(E-\mu)\over T}\right)\right)^{11\over 3}\nonumber\\
&-&{20\times2^{2/3}\over39\pi^{11/3}}\left({T\over m_D}h_6\left({(E-\mu)\over T}\right)\right)^{13\over 3}\Bigg\} ,
\label{diff_t_2}
\end{eqnarray}
where,
\begin{eqnarray}
h_4\left({(E-\mu)\over T}\right)&=&\left[\Gamma(\textstyle 4)
  \left(\Li_{3}(-e^{-\alpha})-\Li_{3}(-e^\alpha)\right)\right]^{1\over 3}\nonumber\\ 
h_5\left({(E-\mu)\over T}\right)&=&\left[\Gamma\left(\textstyle{14\over3}\right)
  \left(\Li_{11/3}(-e^{-\alpha})-\Li_{11/3}(-e^\alpha)\right)\right]^{3/11}\nonumber\\
h_6\left({(E-\mu)\over T}\right)&=&\left[\Gamma\left(\textstyle{16\over3}\right)
  \left(\Li_{13/3}(-e^{-\alpha})-\Li_{13/3}(-e^\alpha)\right)\right]^{3/13}.
\label{exp} 
\end{eqnarray}
After the magnetic part we derive the expression
 of the electric part. In case of electric term one finds,

\begin{eqnarray}
{\cal B}\Big|_l^{soft}(E)-{\cal B}\Big|^{\rm soft}_{l, E=\mu}&=&{e^2m_D^3\over 128}\left({T\over m_D}g_2\left({(E-\mu)\over T}\right)\right)^{4}+O(a^5),
\label{diff_l_1}
\end{eqnarray}
where,
\begin{eqnarray}
g_2\left({(E-\mu)\over T}\right)=[\Gamma(\textstyle 5)
  (\Li_{4}(-e^{-\alpha})-\Li_{4}(-e^\alpha))]^{1\over4}.
\label{exp2}
\end{eqnarray}
Finally, we obtain the expression for longitudinal momentum diffusion-coefficient as,
\begin{eqnarray}
{\cal B}&=&e^2m_D^3\Bigg\{{1\over 72\pi}
\left({T\over m_D}h_4\left({(E-\mu)\over T}\right)\right)^3-{3\times2^{1/3}\over 99\pi^{7/3}}\left({T\over m_D}h_5\left({(E-\mu)\over T}\right)\right)^{11\over 3}\nonumber\\
&-&{20\times2^{2/3}\over 39\pi^{11/3}}\left({T\over m_D}h_6\left({(E-\mu)\over T}\right)\right)^{13\over 3}\Bigg\}
+{e^2m_D^3 \over 128}\left({T\over m_D}g_2\left({(E-\mu)\over T}\right)\right)^{4}.
\label{diff_full_exp}
\end{eqnarray}
This expression is polylogarithmic in nature and also contains fractional power in $|E-\mu|$. This fractional power indicates
 the deviation from Fermi-liquid behavior. This departure can also be seen in the zero temperature case. 
 Hence,
 the final expression for ${\cal B}$ in the extreme zero temperature limit becomes,
\begin{eqnarray}
{\cal B}={e^2 |E-\mu|^3\over 72\pi }-{2^{1/3}e^2m_D^3\over 33\pi^{7/3}}\left({|E-\mu|\over m_D}\right)^{11\over 3}+{e^2|E-\mu|^{4} \over 128m_D}+\ldots.
\label{diff_final}
\end{eqnarray}
The first two terms in the last equation correspond to the transverse contribution and the remaining third term comes 
from the longitudinal interaction. The expression for longitudinal diffusion coefficient has been already obtained in \cite{Sarkar10}. 
Like $\eta$, in ${\cal B}$ also we find that the subleading transverse part is greater than the leading longitudinal contribution.
 We see from the Fig.(\ref{nlo_diff}) that nature of the curve for the diffusion coefficient 
is same as that of $\eta$ as shown in the previous subsection.

\section{Conclusion}

In this paper we have calculated the fermionic drag and diffusion coefficients in a relativistic plasma both
at zero and small temperature by retaining terms beyond the leading contributions. It is seen that the subleading
terms of the transverse sector, which appear with fractional power, are larger than the leading terms coming
from the exchange of soft longitudinal photons or in other words we show that the leading order contributions
to the drag and diffusion coefficients in ultradegenerate plasma cannot be obtained just by adding the leading
order contributions coming from each of these sectors. Both the appearance of the fractional power and dominance of 
the transverse sector are related to absence of the magnetostatic screening or the singular behavior of the 
fermion self-energy near the Fermi surface. Furthermore, we find that the contributions coming from the hard sectors 
are suppressed and the entire physics
is dominated by the soft excitations. This is a clear departure from the finite temperature case where both
the hard and the soft part contribute at the same order. 
As a last remark, we note that here the entire  calculation has been done for QED plasma. It would be interesting to 
extend the present calculation for QCD matter, which might be tricky due to the existence of triple gluon vertex and possible
 magnetic screening in the QCD sector.

\section{Acknowledgments}


S. Sarkar would like to thank P. Roy and S. Chakraborty for their critical reading of the manuscript and T. Mazumdar
 for helpful discussions.


\begin{thebibliography}{50}
\bibitem{Holstein73} T. Holstein, R.~E. Norton and P. Pincus, Phys. Rev. B {\bf 8},  2649  (1973).
\bigskip
\bibitem{Chakravarty95}
S. Chakravarty, R.~E. Norton, and O.~F. Sylju{\aa}sen, Phys. Rev. Lett. {\bf
  74},  1423  (1995).
\bigskip
\bibitem{Boyanovsky01}
D. Boyanovsky and H.~J. de~Vega, Phys. Rev. D {\bf 63},  114028  (2001).

\bigskip
\bibitem{Ipp04}  A. Ipp, A. Gerhold and A. Rebhan,  Phys. Rev. D {\bf 69}, R011901(2004).
\bigskip
\bibitem{Gerhold204}  A. Gerhold, A. Ipp and A. Rebhan,  Phys. Rev. D {\bf 70}, 105015(2004).

\bigskip
\bibitem{Heiselberg93} H. Heiselberg and C. J. Pethick, Phys. Rev. D {\bf 48}, 2916(1993).
\bigskip
\bibitem{Schafer04b}
T.~Sch\"afer and K.~Schwenzer, Phys.\ Rev.\ D {\bf 70}, 114037 (2004).
\bigskip
\bibitem{Pal11}  K. Pal and A. K. Dutt-Mazumder, hep-ph/1101.3870v1(2011).
\bigskip
\bibitem{Brown00}
W.~E. Brown, J.~T. Liu, and H.-c. Ren, Phys. Rev. D {\bf 61},  114012  (2000);
Phys. Rev. D {\bf 62},  054013  (2000).
\bigskip
\bibitem{Wang02}
Q. Wang and D.~H. Rischke, Phys. Rev. D {\bf 65},  054005  (2002).
\bigskip
\bibitem{Gerhold05}  A. Gerhold and A. Rebhan,  Phys. Rev. D {\bf 71}, 085010(2005).
\bigskip
\bibitem{Schafer04}
T.~Sch\"afer and K.~Schwenzer, Phys.\ Rev.\ D {\bf 70}, 054007 (2004).
\bigskip
\bibitem{Boyanovsky01b}
D. Boyanovsky and H.~J. de~Vega, Phys. Rev. D {\bf 63},  034016  (2001).
\bigskip

\bibitem{Bellac97} M. Le Bellac and C. Manuel, Phys. Rev. D {\bf 55}, 3215(1997).
\bigskip
\bibitem{Manuel00} C. Manuel, Phys. Rev. D {\bf 62}, 076009(2000).
\bigskip
\bibitem{Vanderheyden97} B. Vanderheyden and J. Ollitrault, Phys. Rev. D {\bf 56}, 5108(1997).
\bigskip

\bibitem{Braaten91} E. Braaten and T.C. Yuan, Phys. Rev. Lett. {\bf 66} 2183(1991).
\bigskip


\bibitem{Sarkar10}
S. Sarkar and A. K. Dutt-Mazumder, Phys. Rev. D {\bf 82},  056003  (2010).
%







\bigskip
\bibitem{Svetitsky88} B. Svetitsky, Phys. Rev. D {\bf 37}, 2484(1988).
\bigskip
\bibitem{Braaten191} E. Braaten and M.H. Thoma, Phys. Rev. D {\bf 44}, 1298(1991).
\bigskip
\bibitem{Braaten291} E. Braaten and M.H. Thoma, Phys. Rev. D {\bf 44}, R2625(1991).
\bigskip

\bibitem{Mazumder05} A. K. Dutt-Mazumder, Jan-e Alam, P. Roy and B. Sinha, Phys. Rev. D {\bf 71}, 094016(2005).
\bigskip

\bibitem{Moore05} G. D. Moore, D. Teaney, Phys. Rev. C {\bf 71}, 064904(2005).
\bigskip
\bibitem{Roy06} P. Roy, A.K. Dutt-Mazumder, Jan-e Alam, Phys. Rev. C {\bf 73}, 044911(2006).
\bigskip
\bibitem{Mustafa05} M. G. Mustafa, Phys. Rev. C {\bf 72}, 014905(2005).
\bigskip
\bibitem{Peigne108} S. Peigne and A. Peshier, Phys. Rev. D {\bf 77}, 014015(2008).
\bigskip
\bibitem{Peigne208} S. Peigne and A. Peshier, Phys. Rev. D {\bf 77}, 114017(2008).
\bigskip
\bibitem{Beraudo06} A. Beraudo, A. De Pace, W.M. Alberico, A. Molinari, Nucl. Phys. A {\bf 831}, 59(2009).
\bigskip
\bibitem{Vija95}H. Vija and M.H. Thoma, Physics Letters B {\bf 342}, 212-218(1995).
\bigskip
\bibitem{Das10}  S. K. Das, Jan-e Alam, P. Mohanty, B. Sinha,  Phys. Rev. C {\bf 81}, 044912(2010).
\bigskip
\bibitem{Bellac} M. Le Bellac, \textit{Thermal Field Theory} (Cambridge University Press, 1996).

\bigskip
\bibitem{Kapusta} J.I. Kapusta and Charles Gale, \textit{Finite Temperature Field Theory: Principles and Applications} 
(Cambridge University Press, 2006).

\bigskip
\bibitem{Manuel96} C. Manuel, Phys. Rev. D {\bf 53}, 5866(1996).
\bigskip
\bibitem{Pal10}  K. Pal and A. K. Dutt-Mazumder, Phys. Rev. C {\bf 81}, 054906(2010).
\bigskip
 \bibitem{Aurenche99} P. Aurenche, F. Gelis, H. Zaraket and R. Kobes, Phys. Rev. D {\bf 60}, 076002(1999).
\bigskip

 \bibitem{Carrington07} M. E. Carrington, Phys. Rev. D {\bf 75}, 045019(2007).
\bigskip

 \bibitem{Carrington08} M. E. Carrington, A. Gynther and D. Pickering, Phys. Rev. D {\bf 78}, 045018(2008).
\bigskip




\end{thebibliography}
\end{document}